\documentclass{article}

\newcommand{\documentdate}{20 12 2013}
\usepackage{a4wide,graphicx,color}

\title{DRIVING FORCES IN RESEARCHERS MOBILITY}
\author{Floriana Gargiulo and Timoteo Carletti}
\date{\documentdate}

\begin{document}
\begin{titlepage}

\includegraphics[height=3.5cm]{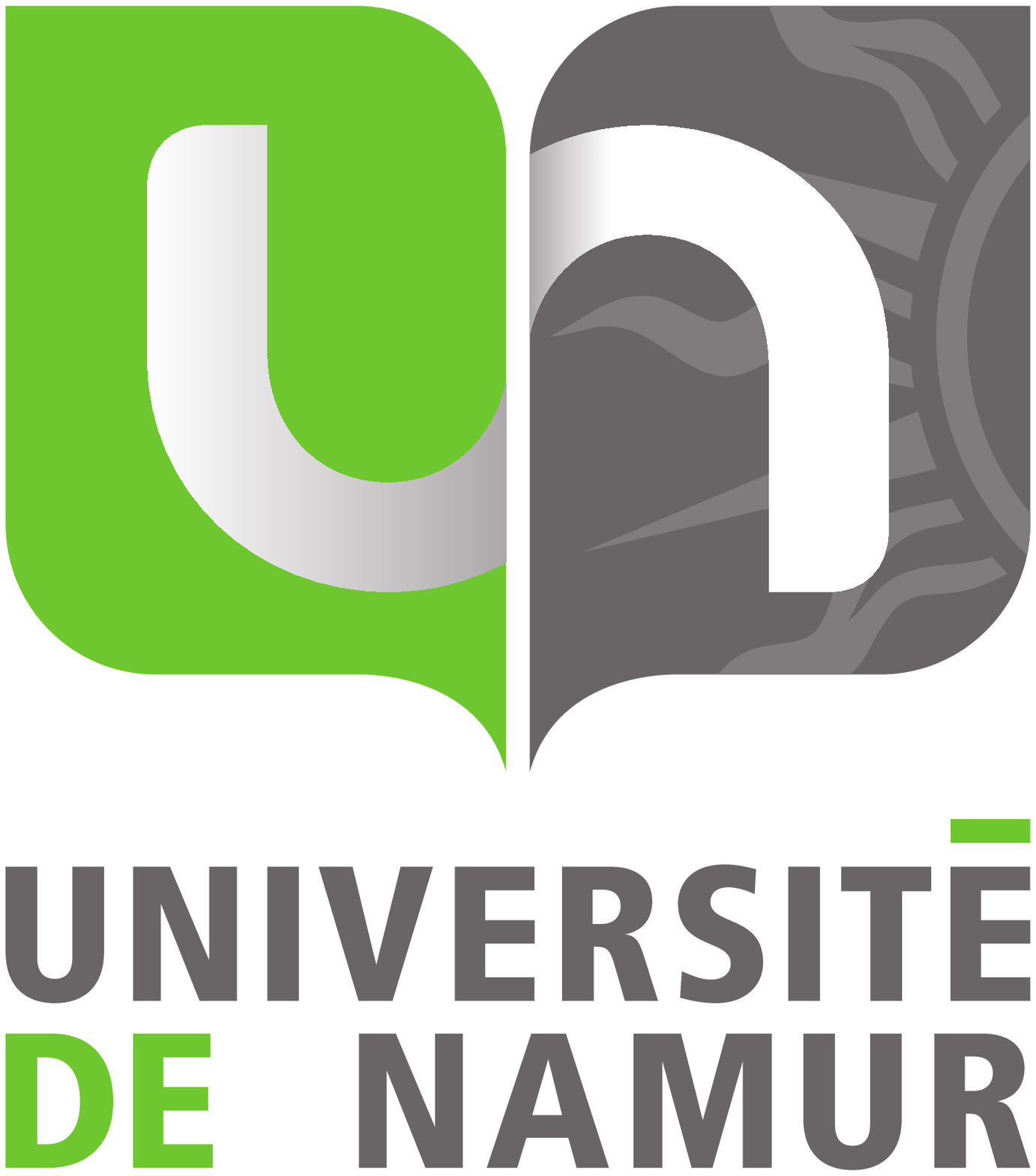}\hfill\includegraphics[height=2.5cm]{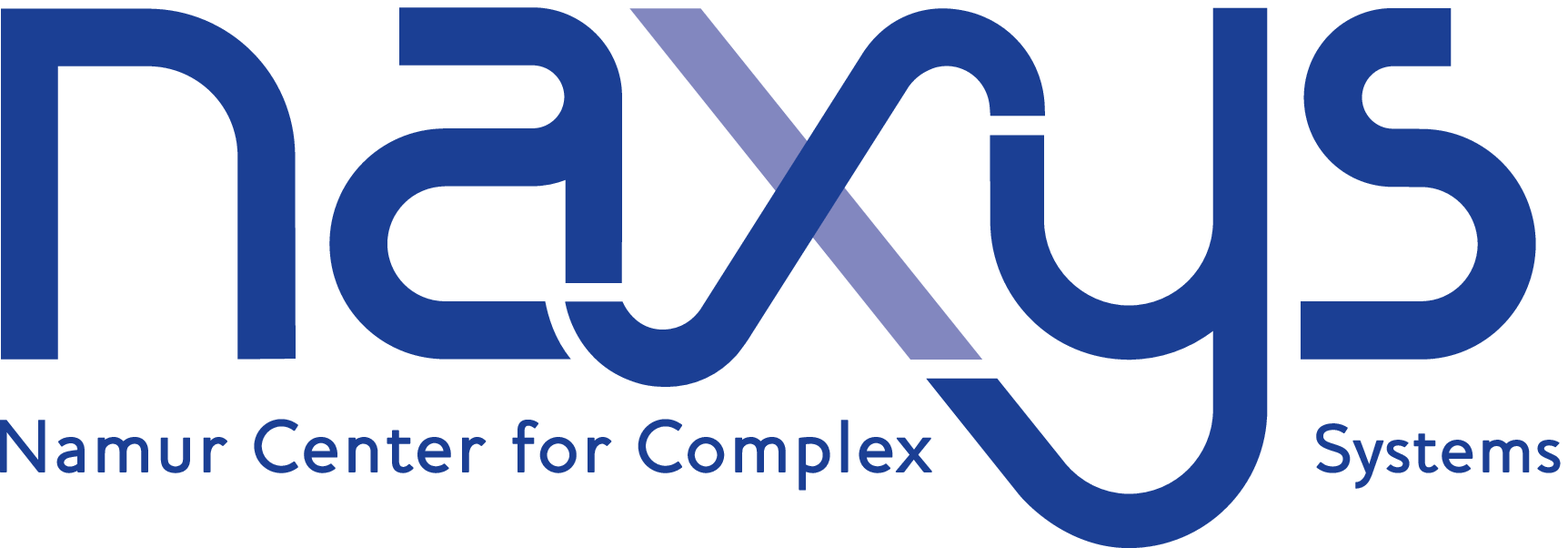}

\vspace*{2cm}
\hspace*{1.3cm}
\fbox{\rule[-3cm]{0cm}{6cm}\begin{minipage}[c]{12cm}
\begin{center}
\vspace{1cm}
Driving forces in researchers mobility\\
\vspace{1cm}
by Floriana Gargiulo and Timoteo Carletti \\
\mbox{}\\
Report naXys-12-2013 \hspace*{20mm} \documentdate \\
\vspace{2cm}
\includegraphics[width=0.6\textwidth]{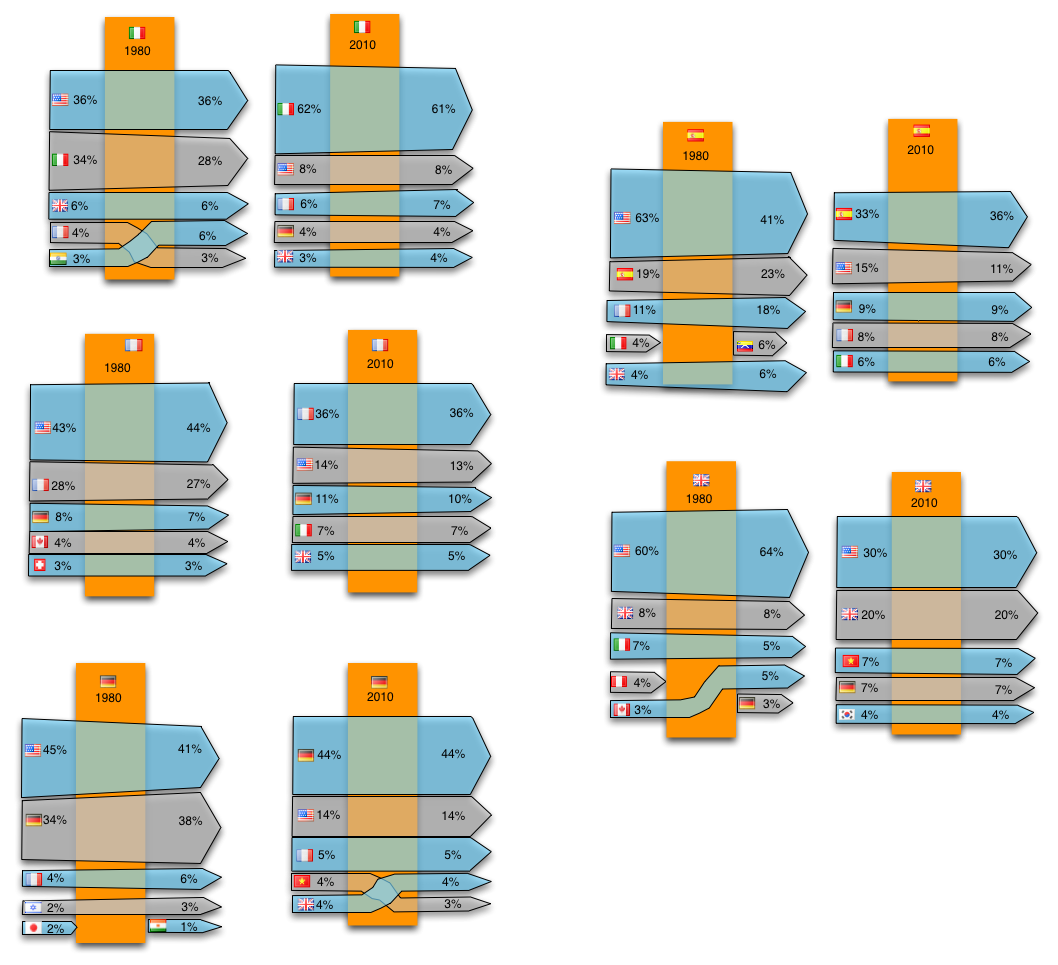}
\mbox{}
\end{center}
\end{minipage}
}

\vspace{2cm}
\begin{center}
{\Large \bf Namur Center for Complex Systems}

{\large
University of Namur\\
8, rempart de la vierge, B5000 Namur (Belgium)\\*[2ex]
{\tt http://www.naxys.be}}

\end{center}

\end{titlepage}
\newpage

\maketitle

\begin{abstract} 
 Starting from the dataset of the publication corpus of the APS during the period 1955--2009, we reconstruct the individual researchers trajectories, namely the list of the consecutive affiliations for each scholar. Crossing this information with different  geographic datasets we embed these trajectories in a spatial framework. Using methods from network theory and complex systems analysis we characterise these patterns in terms of topological network properties and we analyse the dependence of an academic path across different dimensions: the distance between two subsequent positions, the relative importance of the institutions (in terms of number of publications) and some socio--cultural traits. We show that distance is not always a good predictor for the next affiliation while other factors like \lq\lq the previous steps\rq\rq of the career of the researchers (in particular the first position) or the linguistic and historical similarity between two countries can have an important impact.  
 \end{abstract}

\section{Introduction}
Recently a large scientific interest has grown around the analysis of scientific production. In this framework many recent papers have been published concerning "scientometrics", namely the methods to evaluate the scientific production and to forecast the possible success of scientists \cite{scm0},\cite{scm1},\cite{scm2}, \cite{scm3}. This is the reason why, in such papers, the focus has been put on the number of published papers in each individual career.

The idea of using scientific publications in the framework of the complex network science is not new, several studies have been proposed \cite{price},\cite{newman},\cite{jj1}  where the structure and the topological properties of collaborations and citations networks were analysed. In these cases the attention was mainly focused on the citations patterns, the coauthorship communities and on the mechanisms leading to the success of papers in terms of citations. A recent interesting work in this  framework is \cite{fortunato}  where the importance of spatial features in collaborations and citations networks was analysed. The authors showed that geography has a leading role in the morphogenesis of these networks and that the strength of the relationships can be modelled by a gravity law. 

In the present paper, still dealing  with the scientific production data, we focused our research on the mobility in the framework of the research job market. Our aim is to understand which are the driving forces responsible for the choice of an academic position. Some studies on this topic have been developed by the same institutions providing fellowships for international careers (like the Marie Curie) or in the framework of sociological analysis concerning the \emph{talents} migration paths and their impact on social inequalities \cite{rm0},\cite{rm1},\cite{rm2}. In all these cases the analysis was based on surveys, concerning a sample that cannot always be considered to be significant. Moreover, most of these studies were focused to a precise geographical area (mainly at country level), analysing mostly the impact of the local recruitment policies on the academic careers. \\
We use a complex systems approach to analyse the research mobility paths, reconstructing an high number of careers starting from a large corpus of papers published on the journals of the American Physical Society, the spatial trajectory of each author being identified looking at the sequence of its affiliations and their spatial embedding; we hereby assume two subsequent affiliations to be a proxy for the researcher mobility. Let us observe that, while our dataset has the evident limit to be addressed mostly to the physics community, we have on the other hand at our disposal an extremely large number of trajectories to analyse and thus be able to rely our results on good statistics. \\
We first focus on the individual behaviours studying the researchers' trajectories at two levels of granularity: the paths through the universities and the paths through the countries. In this framework a recent literature has been developed concerning human mobility \cite{hm0},\cite{hm1},\cite{hm2}, \cite{hm3}. Even if in this case we can't apply the same statistical tools, given the intrinsic lower length of the researchers paths, we can use a similar formalism concerning the trajectory reconstruction: we study both the trajectories formed by legs regardless of the time duration of each shift between universities and the trajectories where to each leg, the time duration and the beginning of the affiliation period is associated to, therefore the time dimension is fully taken into account in this second case. 
\\Secondly we consider an aggregate scale, focusing on the roles of universities and countries, using a static network approach \cite{reti},\cite{retiW}. We reconstruct the bipartite graph authors - universities (or countries) and its mono--partite projections: the networks of universities and of countries. We analyse the basic topological properties of these graphs and then we study the characteristics of their embedding in the geographical space. Finally we will conclude by studying the impact of the university rank, in terms of publications, on the paths, analysing both the single transitions and the processes with memory. 
\section{The dataset}
The data we use are extracted from the American Physical Society~\cite{aps} publications database containing the records of all the paper issued in the APS journals from 1955 to 2009. Each paper represents one point of the careers of the authors. In such a way, for $N_R=71.246$  researchers in the database, we can reconstruct a trajectory constituted by the temporal sequence of all the distinct affiliations of their papers. The trajectories are embedded in the geographical space using the information extracted by the CEPII geographical database~\cite{cepii} (GIS coordinates for each city, distances between countries, historical and linguistic correlations). More information on the data preprocessing can be found in the \emph{SI}. 
\begin{figure}[hbtp]
\centerline{\includegraphics[width=0.4\textwidth]{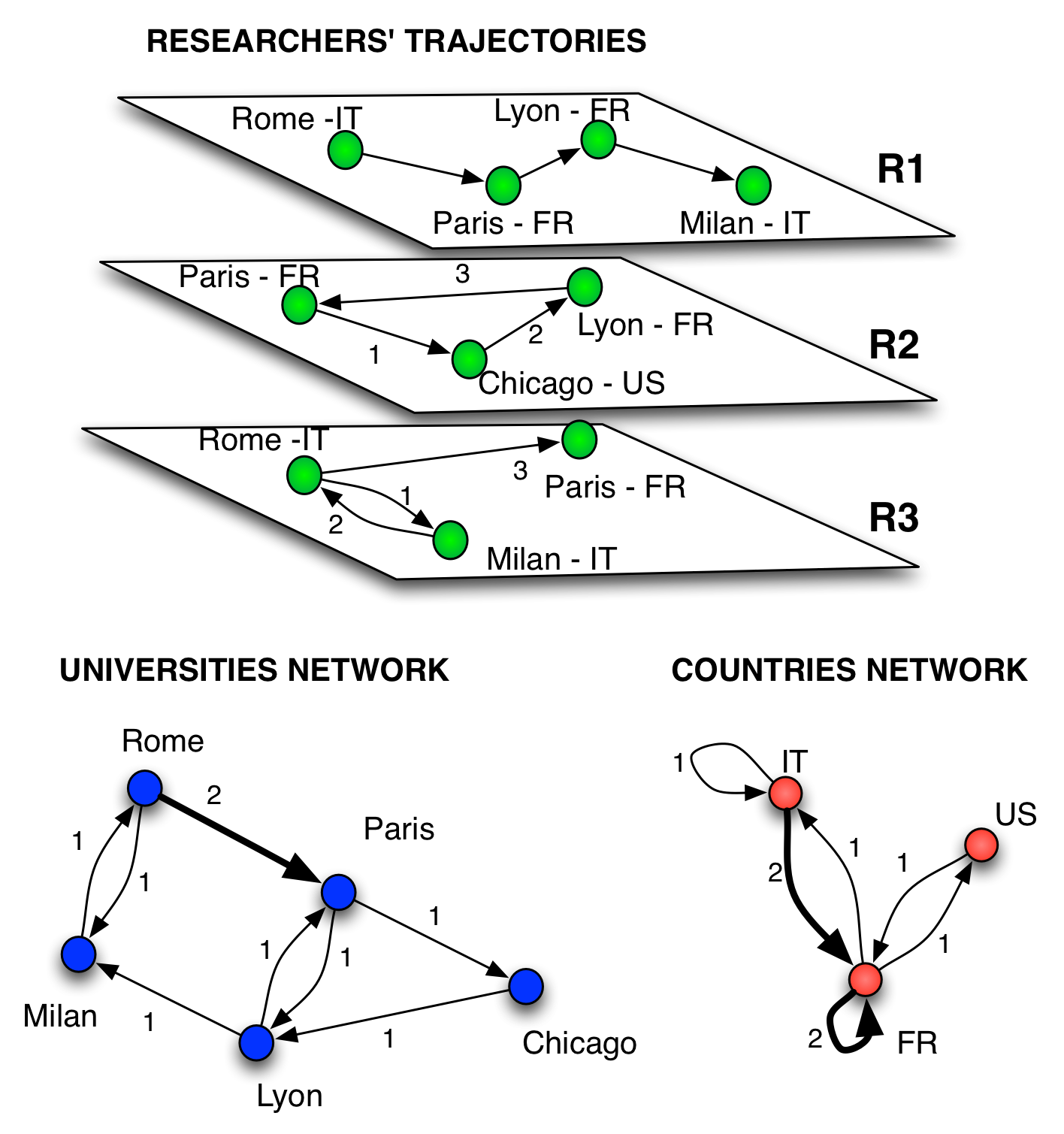}}
\caption{Researchers trajectories, time aggregate universities network and countries network. Top panel: three prototypical researchers trajectories of length three. Each node on a path represents an university the researcher has visited during her/his career, a directed link is drawn between two nodes whenever the researcher moved from one university to a second one, the arrow pointing from the past affiliation to the next one, to denote the time flow direction. R1: four different universities have been visited, R2: the researcher ended her/his career in the same university she/he started visiting thus $3$ universities, R3: after a stay in Milan, the researcher went back to her/his career origin and then moved away. Bottom left panel: the universities network. Nodes denote universities and a directed weighted link is drawn whenever a researcher moved from one university to a second one, the arrow denotes thus the net flow of scholars from one university to another one during the period 1955--2009, the weight being the total number of such moves. Bottom right: the countries network. Nodes represents countries, a directed weighted link between two countries denotes the total number of scholars leaving any university belonging to the first country and going to any university of the second one during the time period under scrutiny.
\label{fig0}}
\end{figure}
All the individual paths can be aggregated in time and then projected to give rise to two possible directed weighted networks (Figure~\ref{fig0}): the universities networks (where the nodes are the $N_{univ}=2434$ universities) and the countries network (the nodes are the $N_{countries}=174$ countries). In both cases two nodes are connected if at least one researcher moves among them (the weight, $w$, of a link is the number of researchers contributing to the link).
\section{Results}
\subsection{Individual level: the researchers paths}
The length $l$ of a path is given by the total number of its legs, even if  the path passes more than once through the same university; this information has not been lost, being captured by studying the total number of distinct countries and universities each researcher has visited, hereby denoted respectively by  $n_{countries}$ and $n_{univ}$. 
We first observe that a large part (more than $40\%$) of the paths have length $l=1$, namely the most probable careers are composed by two academic positions and a single movement among these. This is also confirmed by the distribution of the number of visited universities; on the other side, if we focus on the number of visited countries, we can observe that an important portion of paths just remain in  a single state (see the Figure~1 in the SI). In general we can notice that the $90\%$ of researchers visit at most 4 universities and 3 countries. These results are stable considering careers starting in different time periods (see the Figure~2 in the SI) showing that researchers paths follows prototypical rules that are not evolving in time (PhD+$1$\, postDoc+Permanent position, PhD+Tenure Track, PhD+Postdoc+Leave Academy...). 
\begin{figure}[hbtp]
\centerline{\includegraphics[width=0.6\textwidth]{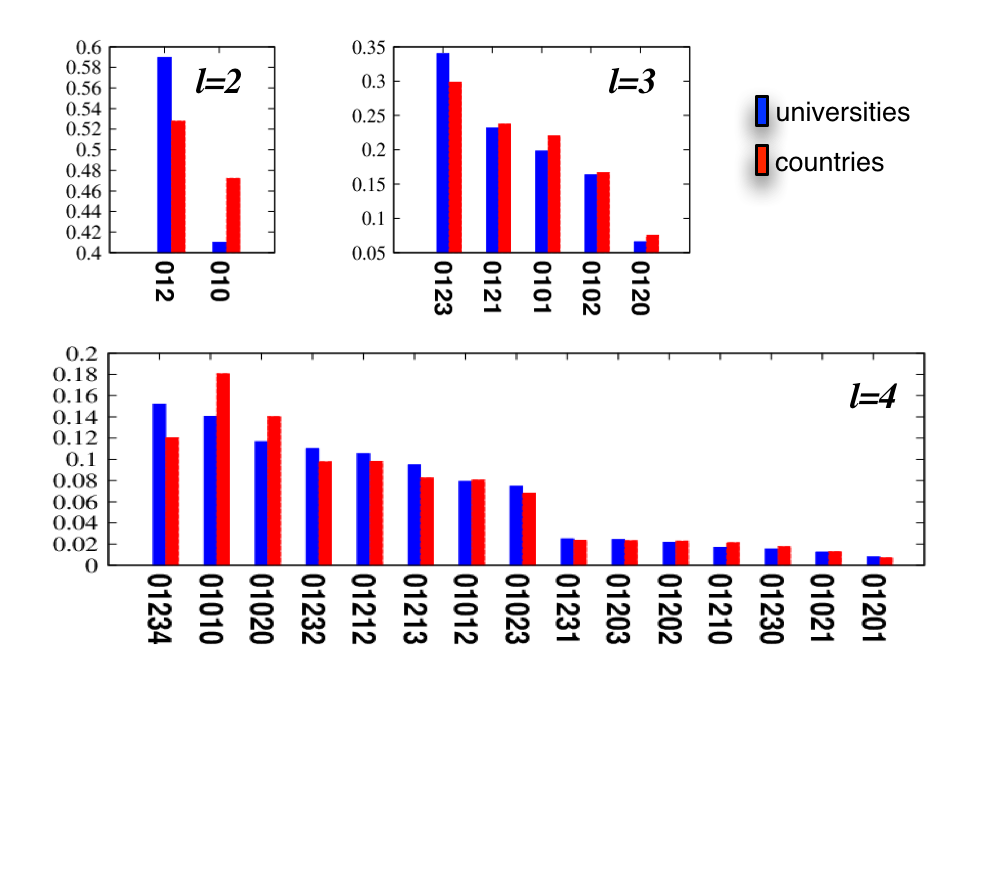}}
\caption{Motifs distribution in the universities and countries networks. Limiting the presentation to paths of length $l=2,3$ and $4$ we show the frequency of the topological structures (linear trees, round trips, 3--cliques, etc.) present in the above networks. From our data we can conclude that researchers with short careers visit mainly different universities, while they tend to remain in the same country showing a propensity to have more than one position in the same state. In the case of longer careers we remark a strong tendency to move back and forth between two universities or two countries, fact that can be associated with a double affiliation of the researchers more than to real mobility.\label{fig1}}
\end{figure}

In Figure~\ref{fig1} we analyse the topological structures, i.e. the exhibited motifs, of the paths (focusing on the paths of length $2$, $3$ and $4$). 
We can observe that for low length paths ($l=2,3$) most of the trajectories connect different universities, paths $012$ and $0123$.  This tendency is a bit less marked at the level of states, showing a certain tendency to have more than one position in the same state, paths $010$ and $0121$ or $0101$).  For longer paths, say $l=3$ and $l=4$, we observe that many patterns are just between two universities, path $01010$. Finally we can notice a very high tendency to return, at some point of the career, to the origin point where the career started, probably corresponding to the place where the personal life of the researcher is centred. These findings have been checked against a null--model. We created a reshuffling algorithm allowing us to preserve the original path lengths distribution and the in and out degrees for each university or country (see SI for a more detailed description). Once applied on our dataset, the new obtained networks will not show anymore such particular motifs distribution, that thus result to be peculiar to the original data. This fact can be quantified computing the metric entropy of each path, both for the real network and for the reshuffled model; even if the most entropic careers ($012$, $0123$, $01234$) are largely present in the real dataset, their frequency is strongly overestimated by the null model. At the same time the reshuffled model is not able at all to reproduce the minimal entropy paths (i.e. the ones connecting just two nodes). The comparison with the base model therefore gives an indication of the fact that the less entropic paths are a typical characteristic of the system (see Figure~4 in the SI).
\begin{figure}[hbtp]
\centerline{\includegraphics[width=0.7\textwidth]{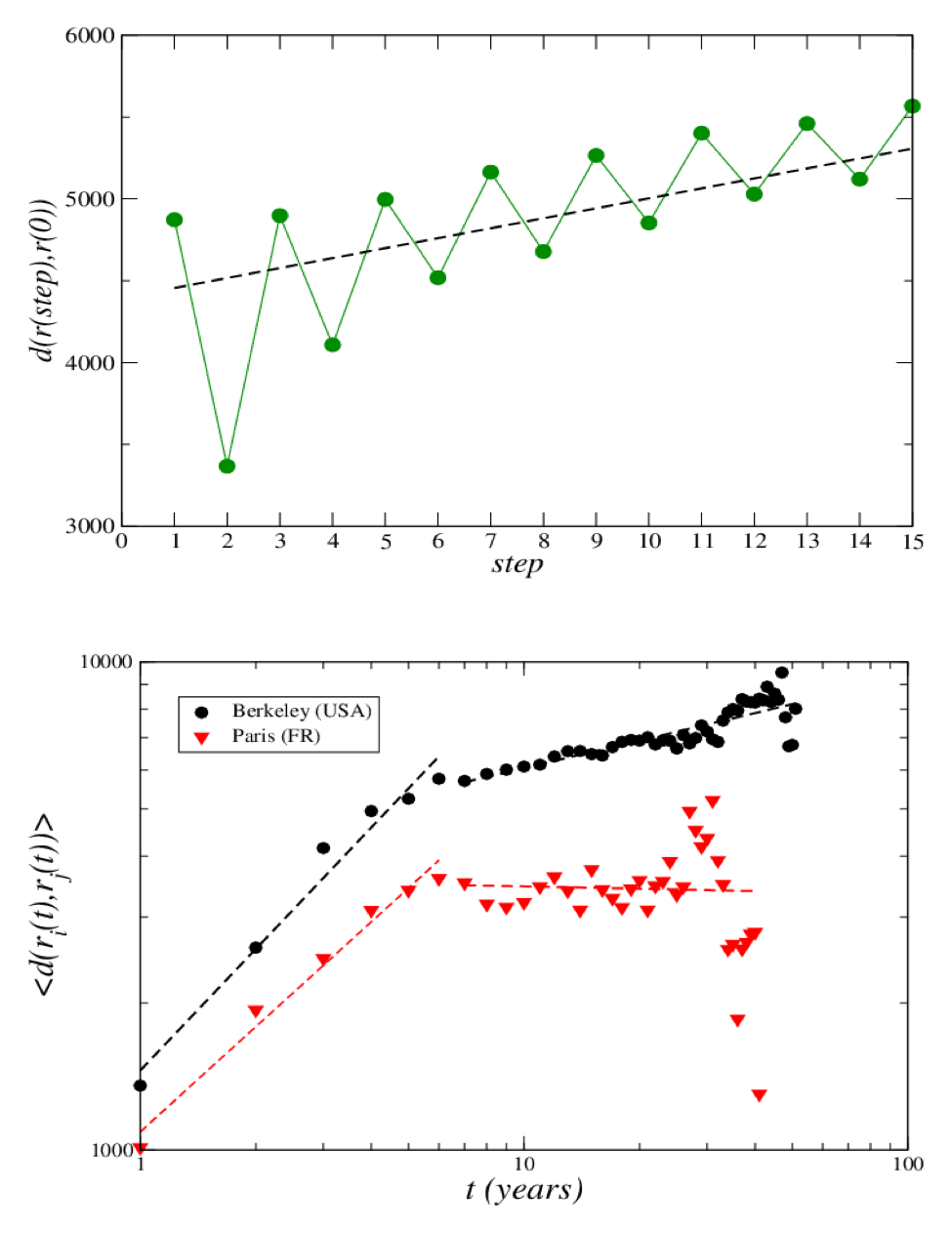}}
\caption{Spatial properties of the average trajectory. We compute the spatial distance between two consecutive affiliations and we studied it as a function of the number of different affiliation and also as a function of the duration of the career measured in years. Upper plot: average distance from the career origin of the successive affiliations, where each trajectory is formed by subsequent steps, regardless of the time duration. The average is calculated for all the trajectories in the dataset. We can clearly see the combination of two antagonistic processes, a diffusive one, tending to push researcher to visit farther and farther universities, coupled with a tendency to get closer to the career origin, probably where the personal life of the researcher is centred. Lower plot: average distance for any couple of trajectories starting from the same career origin as a function of the career duration. For two origins, Paris (Fr) and Berkley (US), we can observe a well defined trend, that we found to be generic across our dataset: initially scholars' trajectories tend to separate far apart quite fast, while in a second phase this behaviour is reduced but still at work (case of Berkley) or even inverted (case of Paris) showing again a tendency to get closer to the career origin in the late phase of the researcher career.\label{fig2}}
\end{figure}
The importance of the first position in the career can be also observed in Figure~\ref{fig2} where the average of all the trajectories considered as sequences of legs regardless of the time duration, shows a gradual spread from the origin with a marked alternated tendency to get closer to the starting point. The lower panel of Figure~\ref{fig2} concerns the trajectories where the time dimension has been taken into account ($t=0$ corresponds to the first paper published with an affiliation, the steps duration correspond to the period of time between two consecutive publications with different affiliations). In this framework, considering couples of trajectories starting from the same position (each one with its own time origin $t=0$) we can analyse their relative distance each year. We report two characteristic behaviours, Paris (Fr) and Berkley (US), and in both cases we observe that after a fast separation during the first $5$ years a stable point is roughly obtained. This suggest that, on average, the time before the stabilisation of the careers is approximately $5$ years. We nevertheless remark two different asymptotic behaviours, careers started at Berkley still continue to diffuse, even if slowly, showing that the US university acts as a career spreader, on the other hand in the case of Paris we observe a tendency to get closer to the career origin in the late years of the academic life.

\subsection{The researchers mobility network}
\begin{figure*}[hbtp]
\centerline{\includegraphics[width=0.7\textwidth]{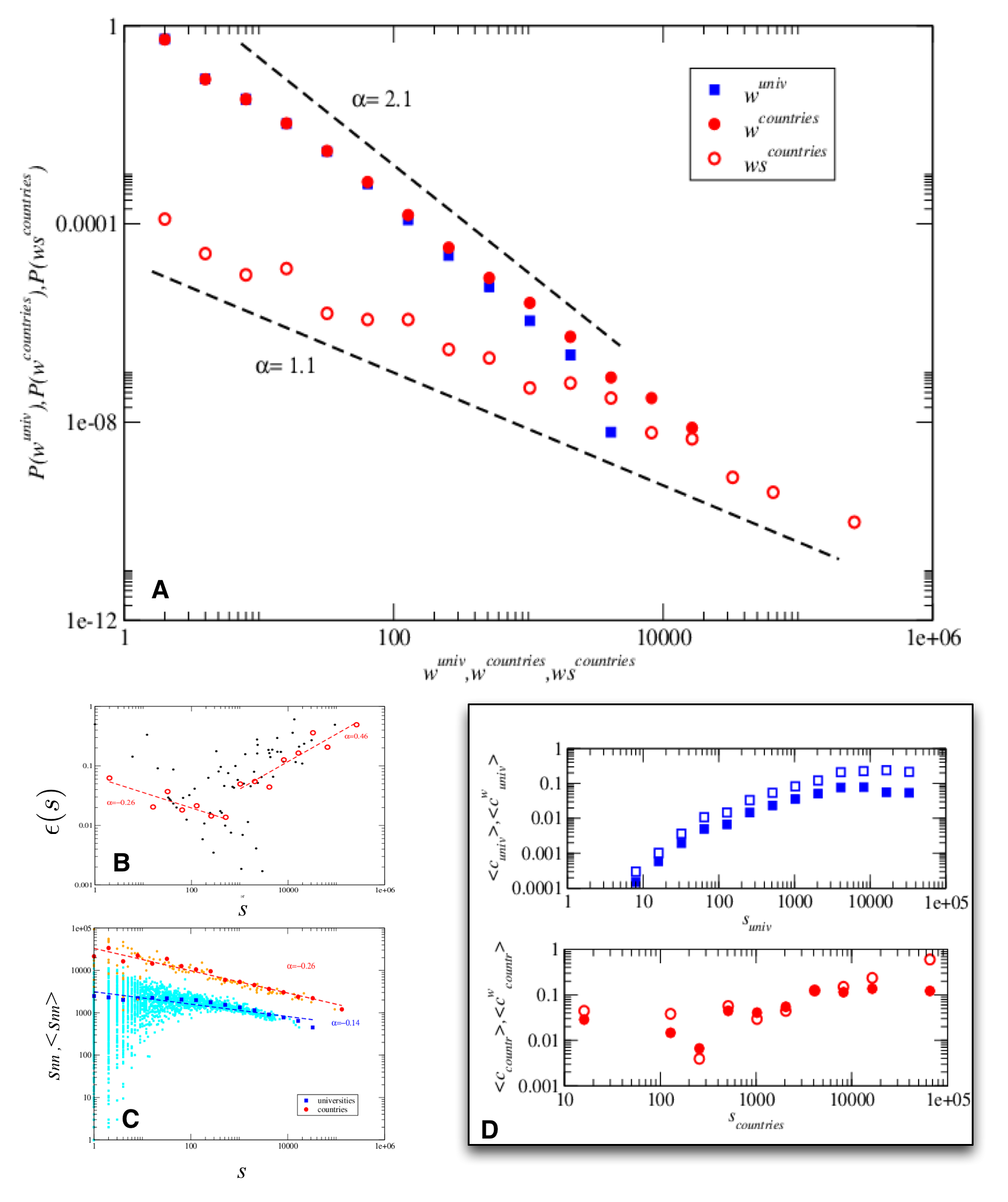}}
\caption{Universities and countries weighted networks. Plot A: Weights distribution for the university network (blue squares), weights distribution for the countries network excluding self--loops (red full circles) and distribution of self--loops weights for countries (red empty circles). All the distributions are broad and display a scale free behaviour. Plot B: Fraction of self--loops with respect to the total number of connections of the nodes as a function of the strengths for the country network. Using the endogamy index, $\epsilon_i=w_{ii}/(s_{tot}^i+w_{ii})$, we can enlighten a non monotone dependence with respect to the total strength, for countries where the international exchanges are very limited there are mainly endogenous movements inducing a decreasing of endogamy as long as few international exchanges are realised; there exists however a tipping point beyond which the more international exchanges are present the more the country acquire international prestige and thus inner scholars have an incentive to rest in the same country, explaining thus the tendency of $\epsilon_i$ to increase with $s_{tot}$. Plot C:  Average total strength of the neighbours as a function of the total strength of the nodes; orange and red circles represent respectively the scattered and averaged values for countries network and cyan and blue squares for universities. One can observe a weak tendency of lower strength nodes to connect to higher degree nodes, that is a dissortative behaviour; however the scatter plot of the individual data shows that low strength nodes are also frequently connected in low strength communities.
Plot D upper panel: Clustering coefficient (full squares) and weighted clustering coefficient (empty squares) as a function of the total strength of the nodes for the universities network. Plot D lower panel: Clustering coefficient (full circles) and weighted clustering coefficient (empty circles) as a function of the total strength of the nodes for the countries network. Because the weighted clustering coefficient is larger than the unweighted one, both for universities and countries, we can conclude that exchanges among the universities or countries are sustained by the weights structures and not only by the topology. Moreover  the clustering coefficients for the universities network, increase with total strength of the nodes, showing the tendency of strong universities to be part of well connected collaboration networks, even if less marked this trend is also present in the case of countries network.\label{fig3}}
\end{figure*}

We start analysing the standard topological quantities of the network structure, both for countries and universities. In the case of countries network self-loops, namely the movements between two universities in the same countries, have an important role and therefore have been considered separately. The self-loops are not taken into account for the calculation of the strengths of the nodes ($s_{in(out)}^i=\sum_{j\neq i}w_{ji(ij)}$, $s_{tot}^i=\sum_{j\neq i}\left(w_{ji}+w_{ij}\right)$).\\
As we can observe in Figure~\ref{fig3}A both for countries and universities, the weights follows a power law distribution, demonstrating a very different connectivity behaviour among couple of nodes. For universities and countries, excluding self--loops the slope of the frequency curve is the same ($P(w)=w^{-2.1}$) while for the self--loops the slope is lower ($P(w_S)=w_S^{-1.1}$). The role of self--loops for each node can be estimated with the \emph{endogamy} index: $\epsilon_i=w_{ii}/(s_{tot}^i+w_{ii})$. As we can observe this index has two different behaviours with respect to the total strength (Figure~\ref{fig3}B): it first decreases with $s_{tot}$ and then after a certain value it shows the opposite tendency. This is due to the fact that some countries, notably not well connected ones or not much active ones, have no sufficient  international impact to exchange researchers  with other countries, and therefore the internal (or at very local scale) mobility is favoured. Of course as the strength increases the international exchanges increase (going outside the country is necessary to increase the personal researchers' impacts) and the endogamy decreases. On the other side, reached a certain level of importance of the countries, researchers could be interested, for linguistic or social reasons, to remain in the same country. The larger is the strength of the country the lower is the effect of the choice of the local mobility on the researchers impacts. This explains the tendency inversion, after a certain value of $s_{tot}$, observed in the figure.   \\
Figure~\ref{fig3}C shows the strength of the first neighbours of a node ($s_{nn}^i=\sum_{j\in\mathcal{V}(i)}s_{tot}^j $) as a function of its total strength. Both for countries and universities there is a weak dissortative tendency (lower strength nodes tend, on average, to connect to higher degree nodes). On the other side, observing the scatter plot of the single measures, we can notice that low strength nodes are also frequently connected in low strength communities. \\ The clustering coefficient (Figure~\ref{fig3}D) is the classical measure of the level of cohesiveness of the neighbourhood of a node~\cite{reti}, $c_i=\frac{1}{2k_i(k_i-1)}\sum_{j,k \neq i}a_{ij}a_{ik}a_{jk}$,
where $a_{ij}$ are the elements of the adjacency matrix of the  unweighted network. If $c_i=0$ the node is in a tree like structure where the neighbours are not connected at all, if $c_i=1$, it is in a completely connected environment. The weighted clustering coefficient is defined as, $c^w_i=\frac{1}{2s_i(k_i-1)}\sum_{j,k \neq i}(w_{ij}+w_{ik})(a_{ij}a_{ik}a_{jk})$.\\
If $c^w_i>c_i$ then the tendency to form triangles is influenced by the weights structures and not only by the topology \cite{retiW}. For the universities network the clustering coefficient increases with the total strength of the nodes, showing a higher tendency for stronger institutions to be inside a fully connected scientific community. The weighted clustering is always higher than the unweighted one, demonstrating that the cliques formation is determined by the weights of the links. For the network of countries we observe again a high clustering for low connected nodes, that is another indicator of the presence of local low degree cliques. 
\begin{figure}[hbtp]
\centerline{\includegraphics[width=0.6\textwidth]{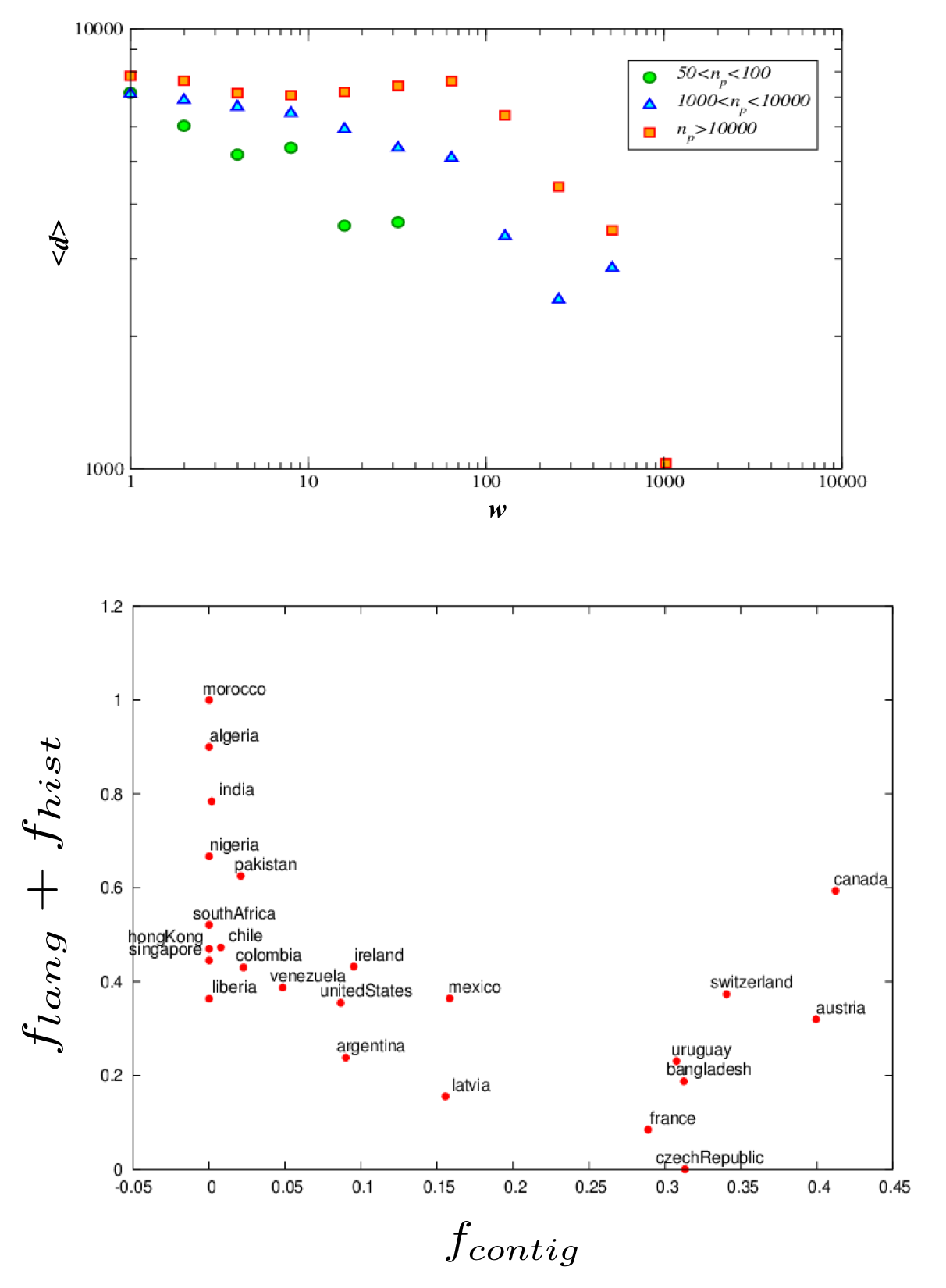}}
\caption{The importance of distance, language and culture. Upper panel: Average distance ($km$) between two nodes as a function of the weights among them for the universities network; green circles represent the low rank universities (number of publications: $50<n_p<100$), blue triangles an intermediary rank ($1000<n_p<10000$), red squares the top rank ($n_p>10000$)). We can observe that scholars' displacements among low ranked universities exhibit correlations with geography similar to the ones presents also in other human movements scenarios. However as long as the rank increases such correlations tend to disappear, showing that distance is not a restraint once researchers move across well reputed universities. Correlations do persist, for high ranked universities, only for very large weights, where even the incentive to reach a good university could not be enough to induce a very large movement. Lower panel, spatial contiguity versus socio--cultural traits:  Each point represents a country with at least $30\%$ of its flows toward either a country with the same language or connected by an historical link either to a contiguous country. On the vertical axis we represent the fraction of flows to a country with the same language ($f_{lang}$) or connected by an historical link ($f_{hist}$). On the horizontal axis we report the fraction of flows to a contiguous country ($f_{contig}$). We can clearly identify two well defined clusters: countries for which the spatial proximity doesn't matter too much, while the exchanges are mainly due to common shared language and/or past history, e.g. Morocco and Algeria with France, India--Pakistan and South Africa with UK, Chile--Colombia and Venezuela with Spain. The second cluster is composed by countries whose flows are mainly directed to neighbours, this is the case of Canada due to its proximity to US, Switzerland and Austria because of their centrality in Europe, but also relatively small countries close to larger ones, e.g. Bangladesh or Uruguay.\label{fig4}}
\end{figure}
In Figure~\ref{fig4} we analyse the correlation between the network morphology and some geo-cultural traits such as the distances, the language and the historical links. In spatial networks \cite{retiS} like mobile phone \cite{hm1,hm2} or commuting networks~\cite{retiComm0, retiComm}, a dependency between the weights of the links and the distance is observed, usually reproducible through a gravity law. This is due to the fact that the choice of the destination is mostly determined by the traveling distance between the points. For researchers, the mobility choices are also influenced by the heterogeneity of the \emph{importance} of the institutions and by the existence of collaboration communities, therefore the direct correlation with the geographic space is partially lost. To explore such phenomenon, we introduce a quantitative indication of the scientific impact of each university, by dividing them into $5$ groups according to their number of publications in the period 1955--2009: (R1) less than $50$ papers, (R2) between $50$ and $100$ papers, (R3) $100$ and $1000$ publications, (R4) $1000$ and $10000$ and (R5) more than $10000$ papers. For the links starting from low rank nodes we can observe that the correlation with geography is maintained: large weights correspond to low distances and viceversa. As the rank increases we can observe that this correlation persists only for extremely large weights. 

Calculating the distance between states is a non-trivial task. Usual measures based on the distances between the capitals didn't allow us to find any correlation between the weights and the distances for the countries network.  Therefore we considered the fraction of fluxes for each node pointing to a contiguous country, to a country speaking the same language or to a country connected by the colonial history.   We found that only the $25\%$ of the countries have a high fraction of fluxes (larger than $30\%$) pointing to a node in one of the previous cases. Therefore we can conclude as a first result that the more obvious drivers for human migrations patterns (geographical and cultural contiguity) are not sufficient to characterise the researchers mobility. For the $25\%$ of countries for which these factors are significant we studied which of the two dimensions (geographical or cultural) is the more important one. In the lower plot of Figure~\ref{fig4}  the cultural dimension, language and historical traits, is represented on the vertical axis and the geographical one on the horizontal axis. In the plane we can identify clusters of countries with similar preferences. For Canada, due to the border with the US, and for Switzerland and Austria (due to centrality in Europe and diffusion of their official languages) both the dimensions are equally important. For other central-European countries, e.g. France, and for small countries contiguous to an important one, e.g. Bangladesh or Uruguay, the geographical dimension is largely prevalent. On the other side we have several cases where the cultural dimension is prevalent, for instance the spanish speaking community, Chile--Colombia--Venezuela, the case of France with the ex--colonies, Algeria and Morocco, and UK with India, Pakistan and South Africa.


\subsection{The importance of being (starting) important}
In this section we study how the choice of a destination country/university depends on the scientific relevance of the country/university in terms of number of publications. 
\begin{figure}[htpb]
\centerline{\includegraphics[width=0.7\textwidth]{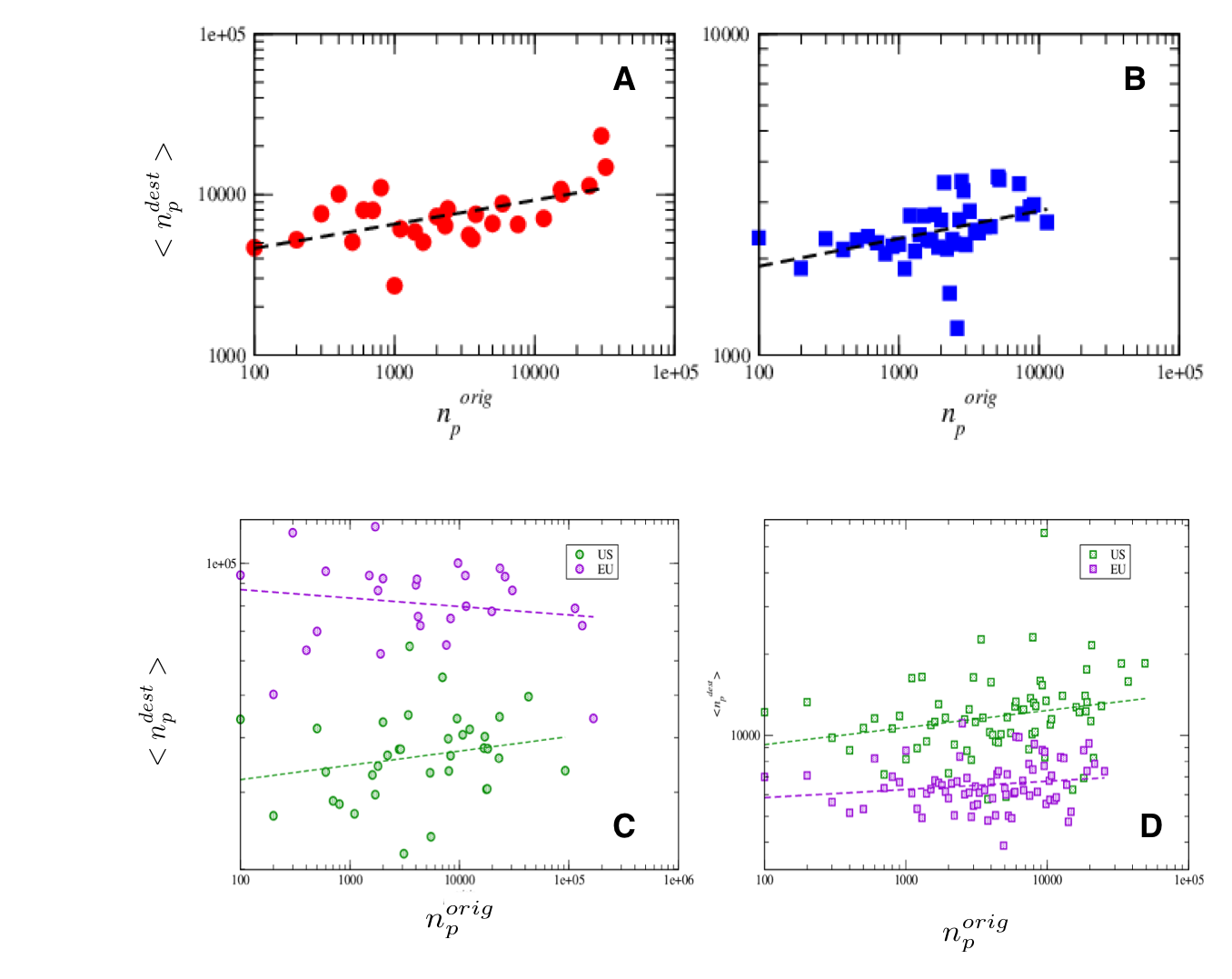}}
\caption{Plot A: Average number of publications of the destination country as a function of the number of publications of the origin one. Plot B: Average number of publications of the destination university as a function of the number of publications of the origin one. In both cases we can observe a weak assortative tendency: universities with a small publication volume tend to be connected, on average, to less important, in terms of publications, nodes and viceversa. However this behaviour is not worldwide spread and differences emerge, for instance Europe versus US, as reported in panels C and D. Plot C: Average number of publications of the destination country as a function of the number of publications of the origin one for European (violet) and US (green) states. Plot D: Average number of publications of the destination university as a function of the number of publications of the origin one for European (violet) and US (green) academies. \label{fig5}}
\end{figure}
In Figure \ref{fig5} we report the results of the study of the correlations between the number of publications in the origin country/university and the number of publications in the destination country/university for each leg of the authors' paths. We can observe that, even if the slope of the regression curve is quite low (much less than a linear regression), an assortative tendency is present both for countries and for universities: less important nodes (in terms of publications) are, on average, connected with less important nodes and viceversa (Figures~\ref{fig5}A-B). Notwithstanding this, these tendencies are quite different for Europe and US, at the level of universities, the assortativity is stronger for US than for Europe. Moreover, at the country level, in Europe we observe a counterintuitive dissortative behaviour. This is due to the fact that inside each European state a large heterogeneity exists at the level of academic institutions; therefore, an excellent university can be surrounded by several other low ranked institutions, lowering the average of the country level (Figure~\ref{fig5}C-D). This point also suggest the fact that at the level of the individual destination choice two different mechanisms can be identified: either a destination institution is selected based on the importance of this institution itself either a country is selected on the basis of other criteria (distance, language, historical links, etc.).

\begin{figure}[htpb]
\centerline{\includegraphics[width=0.8\textwidth]{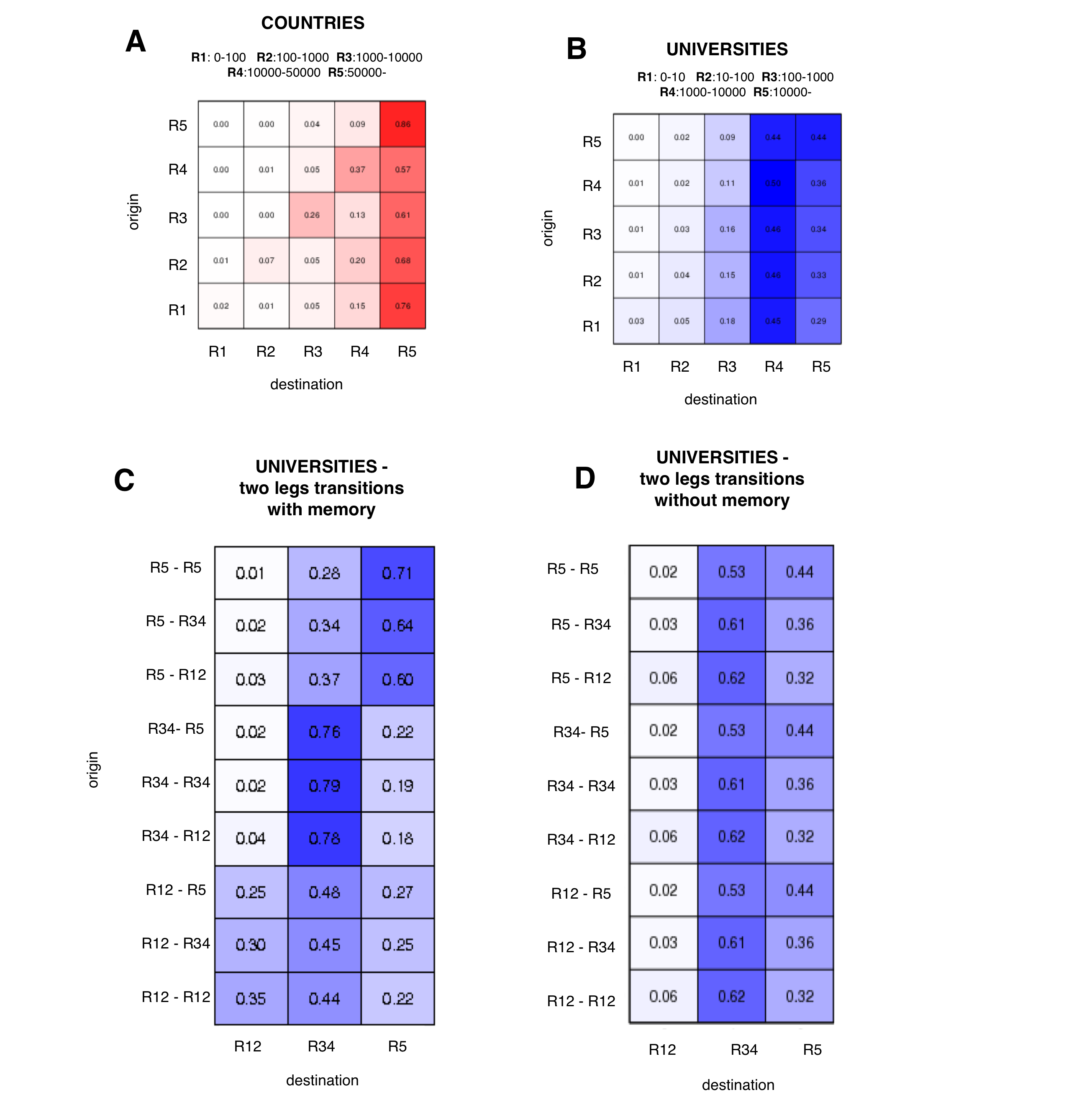}}
\caption{Plot A: probability transitions from each of the $5$ ranked countries to each of the $5$ destinations countries. Plot B: probability transitions from each of the $5$ origins universities to each of the $5$ destination universities. While we can observe that a large amount of movements are toward high ranked countries, in the case of universities most of the scholars trajectories tend to high but not to the highest universities, except for researchers already starting from a highest ranked university. Plot C: transition probabilities with memory. Plot D: transition probabilities without memory. Each value is given by the normalised product of the rank transitions on the two legs. All the possible $2$ legs connections are considered, once we merge the first and the second ranked universities into a unique class R12, and the third and the fourth ones into a single one, R34. From such results we can conclude that the system exhibits a memory effect and thus the first affiliation will play an important role, moreover two consecutive affiliations in a low--medium university will prevent the scholar to reach a higher one, similarly whenever a researcher realises two stays in high ranked universities, her/his career will almost surely remains in the top group universities.\label{fig5b}}
\end{figure}

To deepen the analysis on the career choices according to the scientific relevance of the academies, we resort to the previous division of the countries and the universities into different categories based on the number of published papers and we studied the transition probabilities between the different categories.  At the level of countries (Figure~\ref{fig5b}A), we can observe that a large part of the paths legs points to a high ranked country. In fact, even if these nodes are less numerous, they have the highest number of passing trough traffic. Moreover a sort of homophily between states is observed in the destination choice: many trajectories of the paths connect countries in the same category (notice that the connections between the same country are not considered in this plot). 

At the level of universities (Figure~\ref{fig5b}B), we can observe that in general good institutions (R4) ($1000<n_p<10000$) are the most selected destinations from all the categories. Excellent institutions (R5) ($n_p>10000$) are reached with a higher probability from higher ranked origins.

We also considered a memory (2 legs) process in analogy of what presented in \cite{retiMemory}. Comparing two steps transitions computed from data and the normalised product of the single steps transitions, we can observe that the system exhibits large memory effects and that the probability of reaching a destination is strongly influenced by the first visited institution. Moreover, we can observe that low ranked institutions, that also have on average a very small traffic, are reached with a high probability only from researchers that twice stayed in low level institutions and never by researchers with a past career in better institutions. Also the contrary is true: two consecutive affiliations into high ranked universities will be almost never followed by a low and medium ranked one.

\section{Discussion}
In summary, we analysed the researcher mobility paths from a data driven point of view using the dataset of all the papers published in the American Physics Society journals from 1955 to 2009. We used two different approaches: first we focused on the individual trajectories and then we studied the system at the level of aggregate mobility network structures. The analysis of individual trajectories shows a preference for short paths (between 2 and 4 steps) and that the 90\% of the researchers, in their careers visit at most $4$ universities and $3$ countries. The first point of the path has a particular importance, and the trajectory has an high probability to pass through this point several times. Moreover a sort of memory effect of the starting points is observed when we analyse the time evolution of the average of the distance of all the trajectories from their starting point: an oscillating--like behaviour tending to get closer to the origin point is superposed to a general diffusive dynamics.

The network structures, both at country level and at university level, present heterogeneous weights, degrees and strengths and a general tendency to strength (and degree) assortativity, even if this phenomenon is less pronounced in Europe than in US. Higher degree nodes are usually part of more connected communities (higher clustering coefficient). The role of geography in the network morphogenesis appear to be  relevant only for less important universities. At the level of countries, contiguity seems to be determinant only in particular cases of very geographically central countries, while strong drivers for connectivity are the language and the historical colonial connections. 

In general a large part of the steps of the careers tend to lead the researchers to higher ranked institutions (that are the ones offering most of the positions). On the other side we can observe an important role of the first point of the path: starting from a lower rank institution, independently by the following step, lowers the probability of reaching a top rank academy and makes higher the probability to remain in a low rank one. On the contrary, starting from an high ranked university, the probability of ending in a low rank one is almost zero.

\section*{Acknowledgments}
The authors would like to warmly thank Renaud Lambiotte for useful comments and discussions. \\
This paper presents research results of the Belgian Network DYSCO (Dynamical Systems, Control, and Optimization), funded by the Interuniversity Attraction Poles Programme, initiated by the Belgian State, Science Policy Office. The scientific responsibility rests with its author(s).

\begin{appendix}
\section{Data preprocessing}
The original database contains the information about $457.606$ papers, published on the APS  journals in the period $1955$--$2009$. The XML data structure presented several problems. The main problem, being the lack of one--to--one correspondence between some names of the authors and their affiliations, that could hide homonyms cases. We thus restricted our analysis to a subset of papers free from the above problem.

Secondly also some affiliations were not formatted using the same standard, we hence considered only the cases where the affiliations had the form "laboratory, city, state", in this way we have been able to assign to each author unambiguously, the year of her/his publications, the city and the country where she/he was affiliated at that time. 
After these two filtering procedures the size of the dataset is strongly reduced ($44\%$), leaving $256.890$ publications and $75.358$ authors.

A last problem we were faced to is a typical one arising once using such kind of datasets, that is the disambiguation of the authors, namely to deal with the cases of several authors with the same name. To minimize the impact of the problem we used a cutoff to the length of the paths, considering not realistic a career with more than $15$ affiliations changes. Moreover we checked manually the consistency of all the careers containing more than $10$ legs. This third action reduced the number of authors to $N_R=71.246$. This will be the dataset on which our analysis has been performed.

Starting from such data, the number of publications for each university (or country) is counted directly by the dataset as follows: each paper adds a contribution to all the universities (or countries) to which the $n_{ca}$--coauthors are affiliated. 

Finally we integrated our data with GPS coordinates of each city to which an university was associated to, in our database. The distance between two cities has been calculated using the great circle distance. 

\section{Some features of the researchers careers}
As previously described, for each author we created her/his career path by looking for successive papers with different affiliations, in this way we have been able to associate to each scholar a list of universities - countries - time stamps.

\begin{figure}[h]
\centerline{\includegraphics[width=0.8\textwidth]{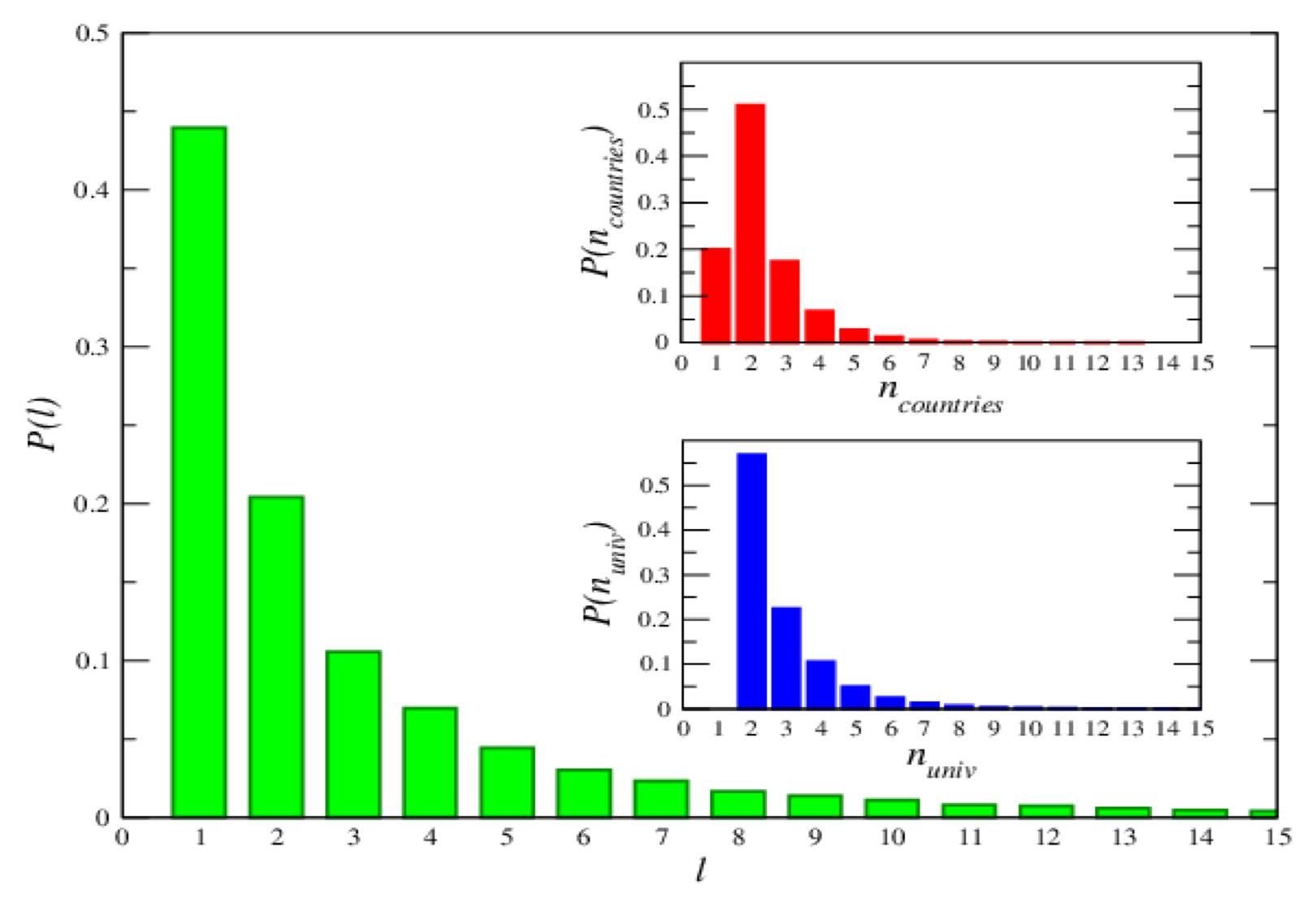}}
\caption{Path length statistics. Frequency distribution of the total length of the path. Upper inbox: Frequency distribution of  the number of visited universities. Lower inbox: Frequency distribution of the number of visited countries }\label{figS0}
\end{figure}

In Figure~\ref{figS0} we report the distributions of total length of the paths, the number of visited countries and visited universities. We can observe that more than $40\%$ of the careers have length $l = 1$, meaning that the most probable careers are composed by just two academic positions and thus a single movement among two universities, as can be straightforwardly confirmed by inspecting the distribution of visited universities, where a clear peak is present for $n_{univ}=2$ (see top inset of Figure~\ref{figS0}). Moreover, our data suggest that is highly probable that, even short careers, will touch (at least) two different countries. In general we can notice that the $90\%$ of researchers visit at most $4$ universities and $3$ countries.

Because our data cover a large temporal span, we have been able to study the historical evolution of the careers duration as a function of the beginning year of the academic life of the researchers. We thus studied some temporal windows of the mobility patterns, namely distinguishing careers starting in different time intervals ($1955$-$1960$, $1960$-$1970$, $1970$-$1980$, $1980$-$1990$, $1990$-$2000$, $2000$-$2009$). To avoid the bias due to the fact that the careers started in the first periods have a longer duration, we just considered for all the cases only the path concerning the first $10$ years of career.

\begin{figure}[h]
\centerline{\includegraphics[width=0.8\textwidth]{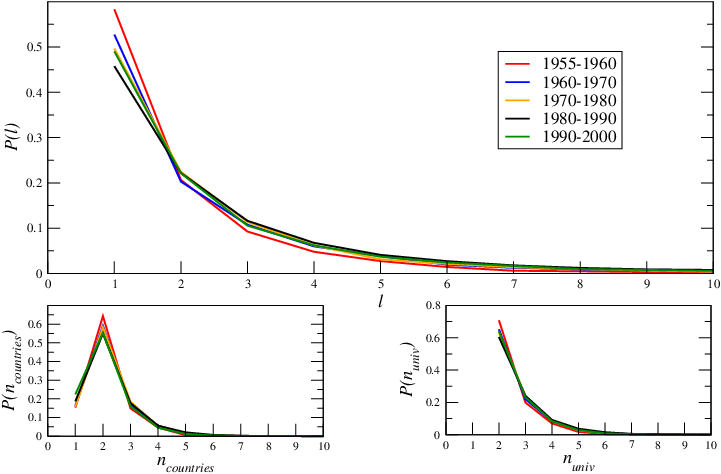}}
\caption{Upper plot: frequency distribution of the total topological length of the first 10 years of career for paths starting in different periods. Lower left plot: frequency distribution of the number of visited countries in the first 10 years of career for paths starting in different periods. Lower right plot: frequency distribution of the number of visited universities in the first 10 years of career for paths starting in different periods. }\label{figS1}
\end{figure}

From the results presented in Figure~\ref{figS1} we can observe that the general shape of the distribution is very stable across the years, we can nevertheless notice that the fraction of short careers ($l=1$) decreases in the last decades, suggesting that the path to a permanent position is becoming longer. At the same time we can observe that the number of researchers visiting only one state decreases, as a confirmation of the increasing of the mobility because of the difficulty to easily find a fixed position.

\begin{figure}[h]
\centerline{\includegraphics[width=0.7\textwidth]{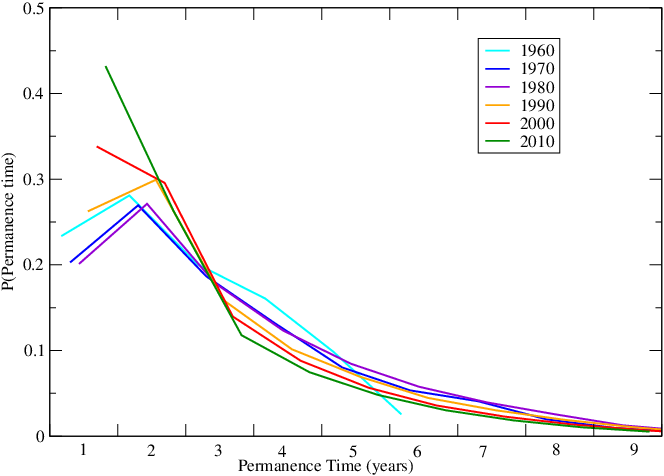}}
\caption{Distribution of the permanence times in each position for paths starting in different periods.}\label{figS2}
\end{figure}

This last finding can be supported by the analysis performed on the time duration of an affiliation in a given university. Iin Figure~\ref{figS2} we report such results and we can clearly observe a strong increase of short duration periods in the last decades, indicating the general increase of mobility in the research careers.

In the main text we presented results supporting the presence of motifs, namely topological structures such as linear trees, round trips, 3-cliques, that are peculiar to the dataset under investigation. To confirm this last statement we built a null--model by applying a reshuffling algorithm that preserves the initial path lengths distribution and the in and out degrees for each university (or country). More precisely starting with all the careers of length $l=1$, we divided them in couples and exchanged the terminations by checking that the so obtained paths are compatible, that is the two universities--nodes in each new career are different. Finally such compatible new careers of length $l=1$ have been used to replace the original ones into the set of all the careers. We then considered all the paths of length smaller or equal to $l=2$ and we repeated the previous algorithm by dividing them into couples and exchanging the terminal nodes. This procedure has been repeated up to careers of length smaller or equal to $l=15$ (the largest possible in our dataset).

To quantify the peculiarity of the motifs we computed for each trajectory $\ell$ the metric entropy
\begin{equation}
\tilde{H}(\ell)=-\frac{1}{\log(|\ell |+1)}\sum_{j \in\ell} p_{j}\log p_{j}\, ,
\end{equation}
where the sum is over all the nodes, i.e. affiliations, in the trajectory $\ell$, $|\ell |$ is the total length of the trajectory and $p_{j}$ is the frequency of the node $j$ in the trajectory. The entropy reaches its maximum value, $\tilde{H}=1$, when all the nodes are visited just once, that is for all nodes $p_{j}=1/(| \ell |+1)$, while it is zero when only one node appears, i.e. there exists a node $\hat{\j}$ such that $p_{\hat{\j}}=1$ and $p_{j}=0$ for all $j\neq \hat{\j}$.

\begin{figure}[h]
\centerline{\includegraphics[width=0.8\textwidth]{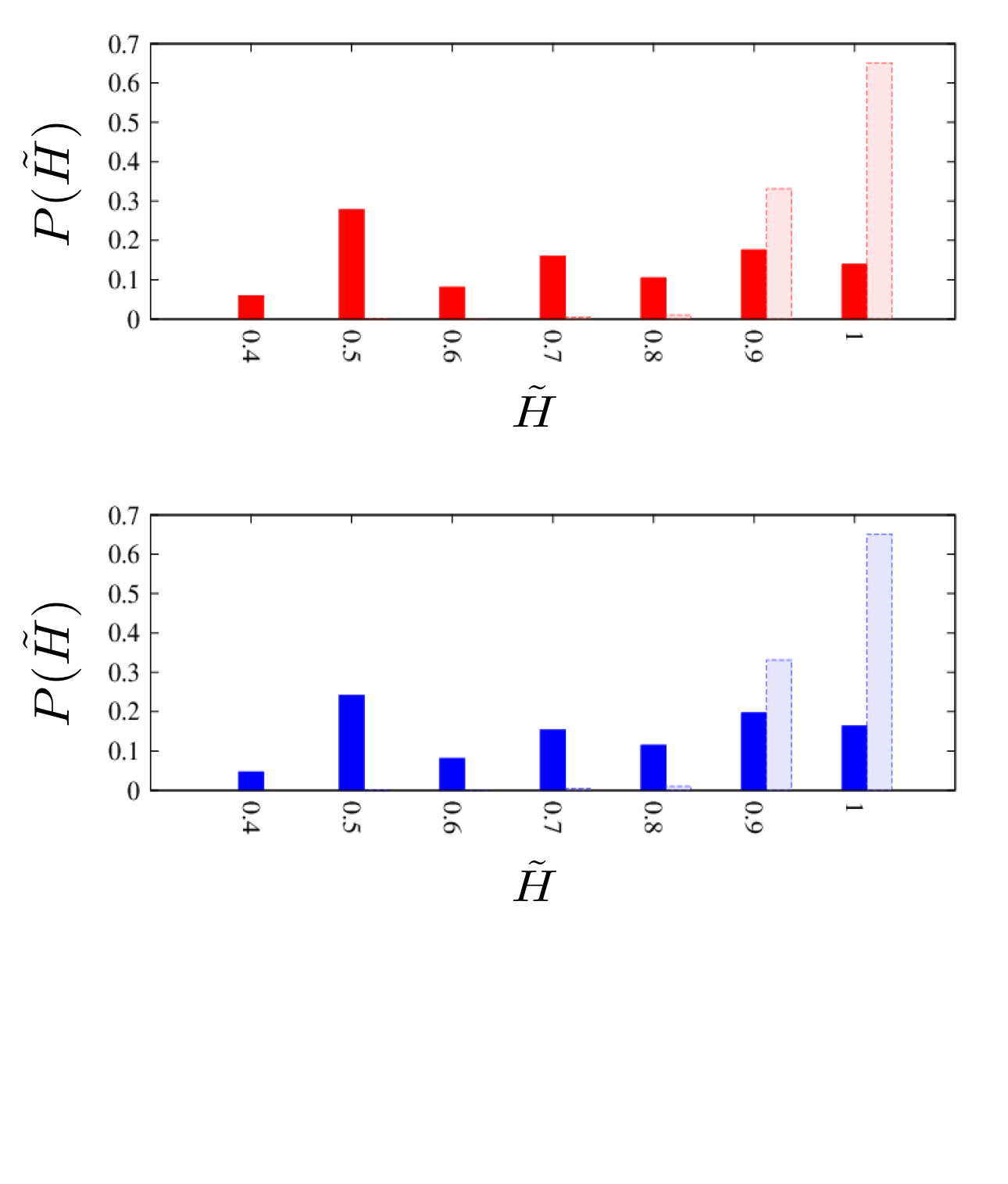}}
\caption{Entropy distributions. Upper plot: countries network. Lower plot: universities network. The full boxes refer to the real data and the empty boxes to the reshuffled model.}\label{figS0_2}
\end{figure}

Results reported in Figure~\ref{figS0_2} clearly show that real data are characterized by an high preference toward low entropy paths, corresponding to frequently pass through a subset of nodes, that are not reproduced at all by the reshuffled model. 

\begin{figure}[h]
\centerline{\includegraphics[width=0.8\textwidth]{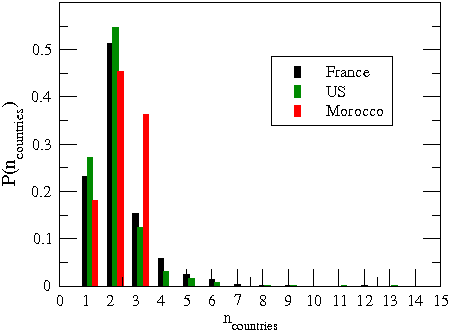}}
\caption{Frequency distribution of the number of visited counties for paths starting in different geographical locations, for three selected countries: France, Morocco and US.}\label{figS3}
\end{figure}

Exploiting the geolocalisation of our data we have been able to emphasise careers differences due to the spatial localisation of the career origin. In Figure~\ref{figS3} we report a study concerning the distribution of the number of visited countries according to the origin state of the career; we can notice that careers starting from European countries, as well as from US, mostly touch two countries, whereas researchers from North Africa countries have a lower tendency to remain in the origin state and, in general focus on two types of paths: either they go to France (or Belgium) and then stop there, either they do a second step to Canada (or US). The same can be observed for the countries in Latin America where the first step is usually to Spain.

\section{Some features of the countries and universities networks}

From the trajectories of each researchers we can construct a network of universities, where nodes are universities and (weighted directed) links among universities denote flow of researchers between the latter ones. Similarly we can aggregate all the universities belonging to the same country to get a countries network.

\begin{figure}[h]
\centerline{\includegraphics[width=1\textwidth]{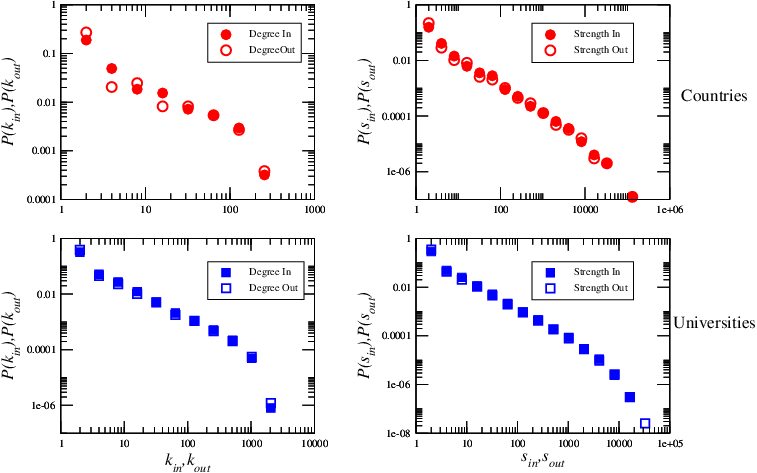}}
\caption{In and out degree distributions and in and out strength distributions. Top panels for the countries network and bottom panels for the universities network. }\label{figS4}
\end{figure}

In Figure~\ref{figS4} we report the in and out degree and strength distributions for the countries and universities networks. For both cases and both for the degree and for the strength, we can observe that the in and out distributions are very similar each other and exhibit a power law distribution.

\begin{figure}[h]
\centerline{\includegraphics[width=1\textwidth]{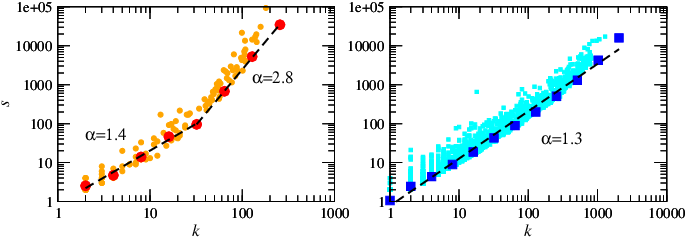}}
\caption{Left Plot: Scattered (orange circles) and average (red circles) strength as a function of the degree of the nodes for the countries network.  Right Plot: Scattered (cyan squares) and average (blue squares) strength as a function of the degree of the nodes for the universities network. }\label{figS5}
\end{figure}

In Figure~\ref{figS5} we plotted the strength of the nodes as a function of their degree. For the universities network we observe an almost linear growth of the strength according to degree. On the other side, for the countries network we can observe two different regimes; the growth is almost linear for low degrees, while it is strongly super--linear for high degrees. This means that, the more a country is able to create research connections, the more it is able to be an important router for the research trajectory.  

\begin{figure}[h]
\centerline{\includegraphics[width=0.8\textwidth]{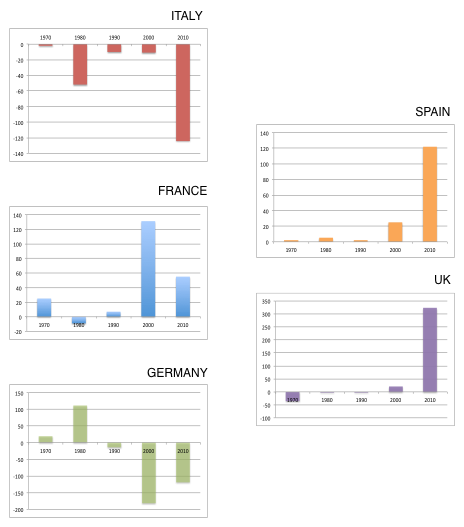}}
\caption{In-flows and out-flows for different European countries (France, UK, Spain, Italy, Germany) in different time ranges:  1960-1970, 1970-1980, 1980-1990, 1990-2000 and 2000-2009}\label{FigS5_1}
\end{figure}

Focusing to the case of some EU countries, we studied the spatio--temporal characteristics of the fluxes in the countries network. In particular we were interested in studying the net flux of researchers leaving and reaching a given country across $5$ temporal windows: $10960$--$1970$, $1970$--$1980$, $1980$--$1990$, $1990$--$2000$ and $2000$--$2010$. In Figure~\ref{FigS5_1} we report the net fluxes ($s_{in}-s_{out}$) for $5$ case studies countries (France, UK, Spain, Italy, Germany). We can notice very different behaviours across countries and in time. For instance, Italy, is characterised as an exporting country during the whole considered period, that is the out flow is always larger than the in flow. On the contrary, France, UK and Spain are strongly attractive for researchers (maybe because of the relative simplicity to obtain long-lasting positions), and this phenomenon is almost constant in time. Germany had a leading role for the immigration of researchers until $1990$ after which the tendency reversed; this can be due to the historical changes that decreased the centrality of Germany in the east-European zone, but also on the policies of researchers recruitment that became much more elaborate. 

\begin{figure}[h]
\centerline{\includegraphics[width=0.8\textwidth]{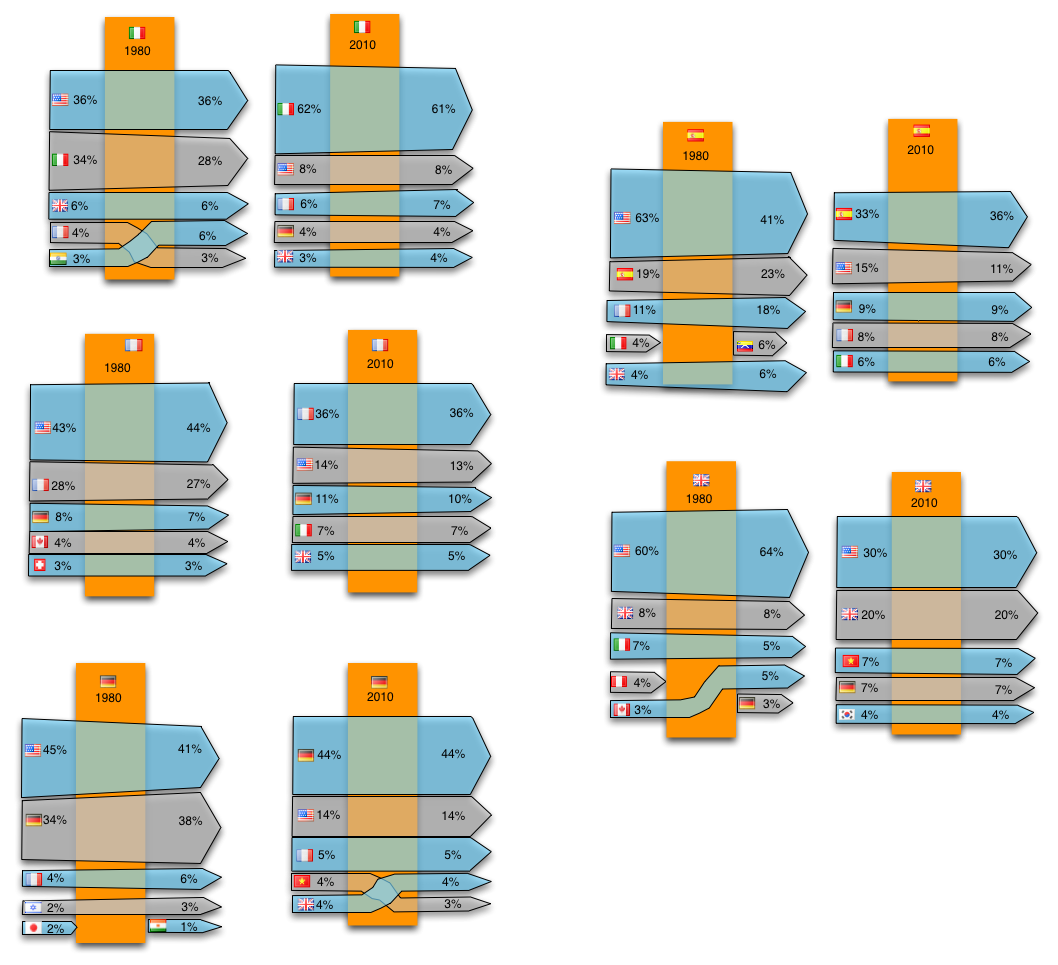}}
\caption{Top 5 flows entering and exiting from five selected European countries (France, UK, Spain, Italy, Germany) in two different periods (1970-1980 and 2000-2009). The left part of the arrow represents the entering flows and the right part the exiting flows. The thickness of the arrow is proportional to the value of the flows.}\label{FigS5_2}
\end{figure}

For each one of the above selected case studies  we also analysed (see Figure~\ref{FigS5_2}) the top 5 immigration and emigration destinations in two different time periods ($1970$--$1980$ and $2000$--$2009$). A first result is the decrease of the central role of US in the international researchers' mobility: in the last decades and in all the case studies countries, but UK - where US remains always the largest exchanger -  we can observe that the largest part of the exchanges are endogamous (toward a different university in the same country). In the former period 1970-1980  we could observe a limited heterogeneity of the international relationships. On one side it was evident the presence of many scientific hubs (US, Japan, Canada), while on the other hand many specialised poles due to historical and cultural connections  are present (Germany - Israel, France - Canada, France - Switzerland, Colombia - Spain). For instance the important relation between Italy and India are due to leading role of the International Center for Theoretical Physics, ICTP, in Trieste founded by the Indian physicist Abdus Salam, strongly active to international exchanges with developing countries. \\

Finally, during the last decade we notice two prototypical behaviors: on one side Italy, France and Spain are consolidating a strong researchers' mobility circuit internal to Europe, while, on the other hand, we can notice strong openings to the emerging Asian scientific authorities (above all China and Korea) concerning Germany and UK. 
\end{appendix}

%
%
%



\end{document}